\documentclass[conference]{IEEEtran}
\IEEEoverridecommandlockouts
\usepackage{cite}
\usepackage{amsmath,amssymb,amsfonts}
\usepackage{algorithmic}
\usepackage{graphicx}
\usepackage{textcomp}
\usepackage{xcolor}

\usepackage{hyperref}
\usepackage{subfigure}
\renewcommand{\baselinestretch}{1}

\begin{document}

\title{RIDE: Real-time Intrusion Detection via Explainable Machine Learning Implemented in a Memristor Hardware Architecture}

\author{
\IEEEauthorblockN{Jingdi Chen}
\IEEEauthorblockA{George Washington University\\
jingdic@gwu.edu}\\   
\IEEEauthorblockN{Gina Adam}
\IEEEauthorblockA{George Washington University\\
ginaadam@gwu.edu}
\and
\IEEEauthorblockN{Lei Zhang}
\IEEEauthorblockA{George Washington University\\
zjiaz\_821117@gwu.edu}\\
\IEEEauthorblockN{Nathaniel D. Bastian}
\IEEEauthorblockA{United States Military Academy\\
nathaniel.bastian@westpoint.edu}
\and
\IEEEauthorblockN{Joseph Riem}
\IEEEauthorblockA{George Washington University\\
joseph.riem@gwu.edu}\\                 
\IEEEauthorblockN{Tian Lan}
\IEEEauthorblockA{George Washington University\\
tlan@gwu.edu}
}



\maketitle


\IEEEpubid{\begin{minipage}{\textwidth}\ \\ \ \\\ \\[12pt]
978-1-6654-2141-6/22/\$31.00~\copyright~2023 IEEE.
\end{minipage}}





\begin{abstract}
Deep Learning (DL) based methods have shown great promise in network intrusion detection by identifying malicious network traffic behavior patterns with high accuracy, but their applications to real-time, packet-level detections in high-speed communication networks are challenging due to the high computation time and resource requirements of Deep Neural Networks (DNNs), as well as lack of explainability. To this end, we propose a packet-level network intrusion detection solution that makes novel use of Recurrent Autoencoders to integrate an arbitrary-length sequence of packets into a more compact joint feature embedding, which is fed into a DNN-based classifier. To enable explainability and support real-time detections at micro-second speed, we further develop a Software-Hardware Co-Design approach to efficiently realize the proposed solution by converting the learned detection policies into decision trees and implementing them using an emerging architecture based on memristor devices. By jointly optimizing associated software and hardware constraints, we show that our approach leads to an extremely efficient, real-time solution with high detection accuracy at the packet level. Evaluation results on real-world datasets (e.g., UNSW and CIC-IDS datasets) demonstrate nearly three-nines detection accuracy with a substantial speedup of nearly four orders of magnitude.
\end{abstract}


\section{Introduction}

Deep learning (DL) has been successfully applied to many network security problems, such as packet inspection and network intrusion detection~\cite{6994301}. However, existing DL-based network intrusion detection algorithms often focus on flow-level features, while large Deep Neural Networks (DNNs) designed for superior detection accuracy have high computation time and resource requirements, making them too heavy to deploy on resource-constrained edge sensors, e.g., in the Internet of Battlefield Things. Real-time, packet-level detection in low-latency, high-speed networks not only requires a substantial speedup over DNNs but also needs 
novel approaches for efficiently combining multiple packet features into a joint signature for detecting sophisticated threats.
Additionally, DNN-based solutions lack explainability, which can be a concern for network operators.

In this paper, we propose an explainable, joint-packet-feature-based solution for network intrusion detection in real time. It consists of three key components. First, the proposed solution makes novel use of Recurrent Autoencoder~\cite{7} to efficiently integrate an arbitrary-length packet sequence into a joint payload-feature embedding for DNN classification. 
Second, to address the limitations of DNNs in terms of explainability for packet-level network intrusion detection while maintaining the superior performance of learning-based approaches, we convert the learned DNN-based classifier into a decision tree policy that is trained by the CART algorithm~\cite{cart} and is inherently explainable. We further employ cost complexity pruning (CCP)~\cite{26} to prune the decision tree for better compactness and ease of implementation. Finally, to support real-time, packet-level detection in low-latency, high-speed network scenarios, the pruned decision tree is then implemented using a flexible memristor architecture. Memristors are novel two-terminal memory devices that are ultra-scalable to a few nanometers and can have their internal state programmed in an analog fashion and retained in an energy-efficient manner~\cite{chen2020reram}. While memristors have been extensively used for the hardware
implementation of neural networks, solutions based on decision trees can also benefit from efficient memristor-based hardware implementations in resource-constrained environments~\cite{ref4}. In this work, we show that our memristor-based analog architecture for network intrusion detection achieves detection speeds of microseconds, together with significant area reduction and energy efficiency. 


We formulate the proposed software-hardware co-design approach as a joint optimization of relevant software and hardware control knobs, (including decision tree pruning parameters and circuit parameters, respectively) with respect to appropriate design constraints. This joint optimization problem is solved using an alternative optimization. 
We evaluate the proposed solution using two widely used open-source network intrusion detection datasets: the UNSW-NB15 and CIC-IDS-2017 datasets. Our experimental results show that our proposed solution outperforms existing DL-based methods 
for both binary- and multi-class classification tasks, achieving substantial speed-ups and providing an explainable and real-time learning-based solution specifically tailored for high-speed networks with low latency requirements.

Our main contributions are as follows:
\begin{itemize}
    \item We propose a novel algorithm leveraging autoencoder to integrate multiple packets with a variable number of payload features into joint embeddings for packet-level intrusion detection.
    
    \item We develop a tailored algorithm suitable for hardware mapping that can provide real-time, packet-level network intrusion detection in constrained computational environments while maintaining the detection performance.

    \item We propose the co-exploration and co-development of emerging hardware and algorithmic technologies to create techniques of future use in autonomous cyber operations, which combines insights into cybersecurity and machine learning with advances in nanoscale devices to co-design and prototype a memristor hardware-mappable algorithm for learning-based network intrusion detection. 
    
    \item Evaluations show that our design significantly outperforms DL-based intrusion detection algorithms, achieving nearly three-nines detection accuracy and a substantial speedup of nearly four orders of magnitude with the memristor-based hardware implementation.
\end{itemize}


\section{Related Works and Background}
\subsection{NIDS for IoT systems based on Learning Techniques}

Network intrusion detection systems (NIDS) have become increasingly important in network security, and various machine learning methods, including supervised, semi-supervised, and unsupervised learning, have been used to enhance their accuracy and precision in detecting anomalies. Several studies have proposed NIDS for IoT networks using different machine learning algorithms and architectures, such as Multilayer Perceptron(MLP)~\cite{hodo2016threat}, Artificial Immune System(AIS)~\cite{hosseinpour2016intrusion}, 
Internet of Things Intrusion Detection and Mitigation(IoT-IDM)~\cite{nobakht2016host}, 
and Conditional Variational Autoencoder(CVAE)~\cite{lopez2017conditional}. Other studies have recommended the use of fog computing to improve efficiency and scalability in IoT systems~\cite{diro2018distributed}.
However, previous DL-based algorithms in NIDS have limitations, such as failing to effectively analyze and coalesce features from a sequence of packets, being too heavy to deploy in resource-constrained environments, and lacking explainability. Additionally, real-time detection in high-speed networks requires a significant speedup over DNNs, which may use millions to billions of parameters, making them unsuitable for packet-level intrusion detection in low-latency, high-speed network environments.

\subsection{Implementing Decision Tree with Memristors}

From a circuit design perspective, keeping the size of the decision tree tractable, high reusability, and flexibility of the blocks is important to compute in a resource-constrained environment. A pruned decision tree with fewer nodes is friendlier for a hardware translation since some of the sub-trees of the decision are replaced with leaves. 
One of the more mature emerging device technologies for edge computing is the memristor~\cite{akinaga2010resistive}, which is a good candidate for this endeavor since a fully analog circuit with high efficiency can be implemented. For instance, a single molecular memristor with complex engineered functionality has successfully implemented a decision tree with 71 nodes, utilizing voltage-driven conditional logic interconnectivity among five distinct molecular redox states~\cite{goswami2021decision}. Another example involves an analog front-end design where a memristor device serves as an adjustable boundary, facilitating the comparison of a programmed boundary with input data~\cite{ref4}. Furthermore, decision trees can be implemented in analog content addressable memories based on memristor devices~\cite{li2020analog}. The development of memristor devices paves the way for low-energy, high-speed, and distributed in-memory computing architectures for decision-making in hardware.

\section{Problem Formulation}
\label{sec: problem formualtion}


The problem of network intrusion detection can be formulated as a predictive classification modeling problem. Let $\boldsymbol{X}$ represent the set of input variables or features, which could include characteristics such as packet size, source, and destination addresses, protocol type, etc. Let $\boldsymbol{Y}$ represent the set of output variables, where $y \in \boldsymbol{Y}$ corresponds to the classification label of a record, indicating whether it is normal or an intrusion. Let $D = \{(x_1, y_1), (x_2, y_2), ..., (x_n, y_n)\}$ denote the training dataset, i.e., the UNSW-NB15 dataset or CIC-IDS-2017 dataset, where $x_i \in \boldsymbol{X}$ represents the feature vector of the i-th record, and $y_i \in \boldsymbol{Y}$ represents its corresponding classification label.
Let $f: \boldsymbol{X} \to \boldsymbol{Y}$ represent the classification function or classifier that maps an input feature vector $x$ to its corresponding classification label $y$.
Since the flow-level features are insufficient for intrusion detection, we need to harness packet-level features for the training datasets. 
Define a sequence of packets to be $\xi = [p_1,p_2,\dots,p_N]$, where $p_i$ represents packet $i$ and gives values to input feature vectors $x_i$ for training. Note that packets have different lengths. 
Denote the hardware configuration $\mathcal{H}$ to be the memristor-based solution approximating classification function $f$. Thus, we can express the learning classification accuracy as $\hat{\Phi}(\mathcal{H}, \xi)$.
To optimize the learning accuracy with hardware configuration $\mathcal{H}$, we define a function $\mathbb{E}_{\xi \in D} [\textbf{1}\{\hat{\Phi}({\mathcal{H},\xi}) = \Phi(\xi)\}] $ that is based on the expectation over the evaluating datasets $D$, where $\textbf{1}{\cdot}$ denotes the indicator function, $\Phi(\xi)$ is the learning accuracy without hardware implementation. 
Our goal is to identify the optimal configuration $\mathcal{H}$, which is the combination of both software and hardware side that maximizes the expected accuracy $\mathbb{E}_{\xi \in D} [\textbf{1}\{\hat{\Phi}({\mathcal{H},\xi}) = \Phi(\xi)\}] $. The resource constraints of the solution $\mathcal{H}$ are under consideration, which is $\zeta (\mathcal{H}) \le Z_{\rm max}$, where $\zeta (\mathcal{H})$ is the overall resource consumption (i.e., power consumption, circuit area consumption) of the hardware solution $\mathcal{H}$, $Z_{\rm max}$ is the resource constraints.

This optimization problem faces two challenges. Firstly, an effective method is needed to compress and integrate packet features into joint embeddings due to the varying lengths of packets $p_i \in \xi$. Secondly, the black-box nature of the DL-based intrusion detection algorithm makes it too heavyweight to deploy in hardware and lacks explainability. Thus, there is a need to develop a tailored, explainable algorithm that is suitable for hardware mapping while maintaining the effectiveness of learning-based approaches.

\section{Methodology}


\subsection{Solution Overview}

\begin{figure}[b]
\centering
\includegraphics[width=3.55in, height=1.8in]{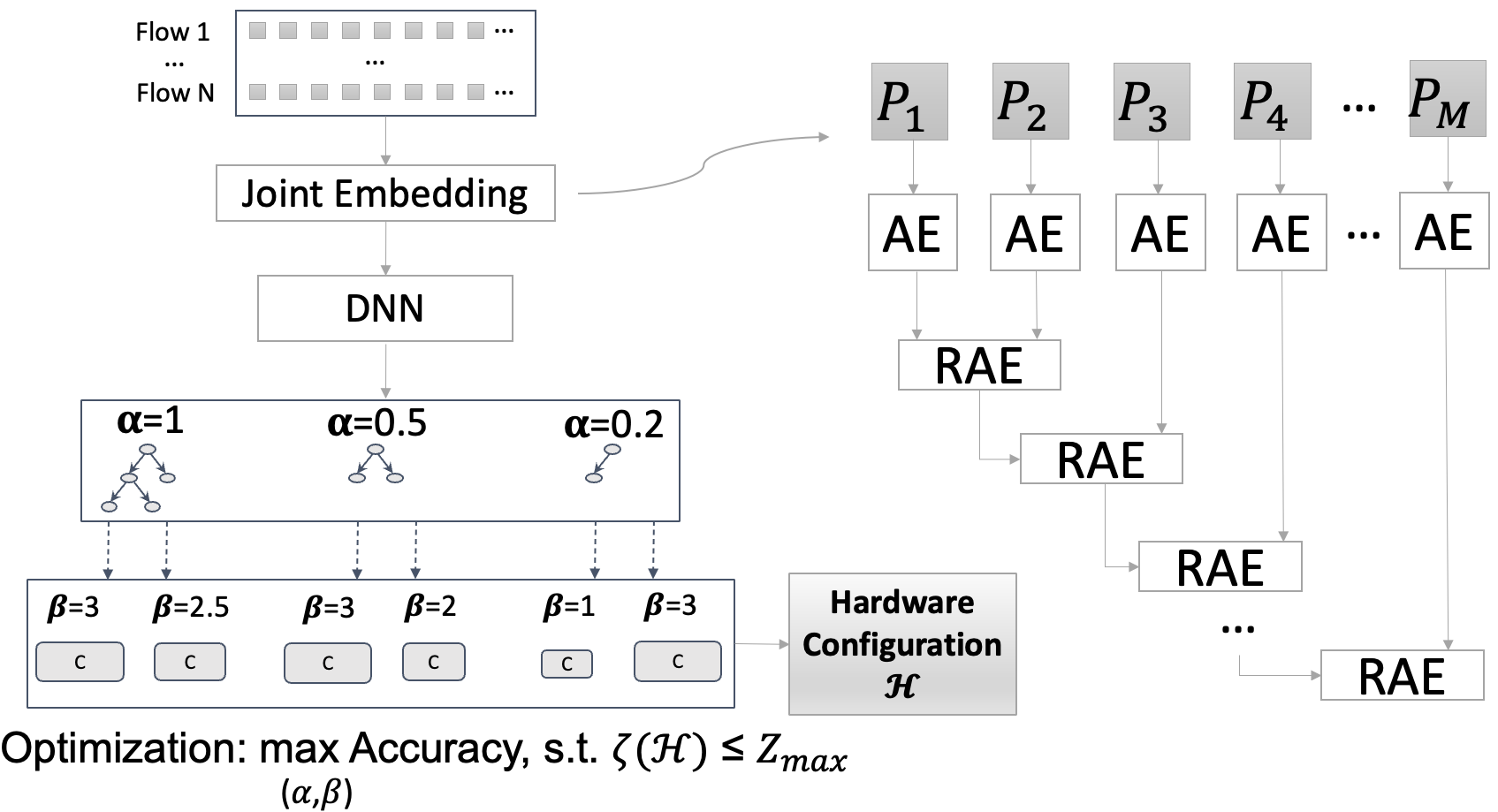}
\caption{Solution Framework}
\label{fig:sys_diag}
\end{figure}

Our design is illustrated in Fig.~\ref{fig:sys_diag}.
To obtain more compact packet joint embeddings, we adopt a greedy method~\cite{white2016deep} to train the Autoencoder and recursively combine pairwise packet feature embeddings. The greedy method uses a hierarchical approach - it first combines feature embeddings of adjacent terms in each connection statistics group and then combines the results from a sequence of connection groups in an execution path. 
Then, for the proper choice of DNN, we train a DNN classifier parameterized by $\theta$(represents the set of parameters or weights of a DNN) for intrusion detection with the use of these new packet joint embeddings. 
Next, we use a teacher-student training methodology~\cite{7} that converts the learned DNN classifier into a decision tree policy that is explainable, then prunes the extracted decision tree $\mathcal{T} = F_{\rm T}(\theta, \alpha)$ with an adjustable decision tree pruning parameter $\alpha$ and implements it using a flexible memristor architecture parameterized by $\beta$(represents the set of parameters of circuit design, we use $c$ in Fig.~\ref{fig:sys_diag} to represent the circuits area overhead, larger $\beta$ corresponds to larger circuit area $c$), whose performance is a function of the extracted decision tree $\mathcal{T}$ and $\beta$, which means the performance metrics of the decision tree implemented by memristor is $ \mathcal{H} = F_{\rm H} (\mathcal{T}, \beta)$. 
Now the accuracy of learning $\hat{\Phi}(\mathcal{H}, \xi)$ can potentially depend on both the performance metrics of the decision tree implemented by memristor $\mathcal{H}$ and the testing sequence of packets $\xi = [p_1,p_2,\dots,p_N] \in D$.

To this end, our objective is to identify the optimal combination of $\alpha$ and $\beta$ that maximizes the expected accuracy. Therefore, the optimization problem in Section~\ref{sec: problem formualtion} can be re-written as a \textit{Joint-Software-Hardware Co-design (JSHC)} optimization problem:
\begin{eqnarray}
  & \max_{\alpha, \beta} &  \mathbb{E}_{\xi \in D} [\textbf{1}\{\hat{\Phi}({\mathcal{H},\xi}) = \Phi(\xi)\}] ; \label{eq:0} \\
& {\rm s.t.} & \mathcal{T} = F_{\rm T} (\theta, \alpha); \label{eq:F_T}\\
& & \mathcal{H} = F_{\rm H} (\mathcal{T}, \beta); \label{eq:F_H}\\
& & \zeta (\mathcal{H}) \le Z_{\rm max} \label{eq:constraints}.  
\end{eqnarray}
The co-design of software and hardware is demonstrated in the lower part of Fig.~\ref{fig:sys_diag}. A larger value of $\alpha$ results in a decision tree with more nodes, then for each fixed value of $\alpha$, we implement the corresponding decision tree with different memristor architectures that were parameterized by $\beta$. During the co-optimization stage, we tune one parameter at a time among $\alpha$ and $\beta$, iteratively tuning them to obtain the highest accuracy score. In the following sections, we will introduce each step of the solution in detail.

\subsection{Integrating Multiple Packet Features for Joint Embedding}
\label{subsec:pac_joint_embeddings_generation}



This section presents a method to extract and compress single packet features to create more compact feature embeddings. We then integrate an arbitrary number of packet feature embeddings into a joint embedding, which serves as the input for the DNN classifier used for intrusion detection prediction. Each packet joint embedding contains features from a variable number of packets, and each packet also has payloads with different lengths, which makes feature extraction more difficult. Our method is the first proposal to address this challenge.

\textbf{Extracting the Packet Features: }
For each packet, $p_i$ in the testing sequence of packets to be $\xi = [p_1,p_2,\dots,p_N] \in \mathbb{R}^{N\times F}$,  where $N$ is the number of packet samples and $F$ is the number of features in each sample, we used a packet extraction method introduced in~\cite{payloadbyte} to analyze packet capture (PCAP) files and extract relevant information. We structure the feature vector for each packet by capturing raw bytes from the packet data and extracting features from the packet header using the parser module $f_p$. To avoid potential overflow or truncation issues, a payload range of $N_p$ bytes was established, ensuring the incorporation of every byte. Each byte was then converted from its hex value to an integer ranging between 0 and 255, resulting in $N_p$ features. For cases with less than $N_p$ payload bytes, zero padding was applied to maintain a consistent feature vector structure. Thus, The raw payload $l_i$ in each packet $p_i$ is transformed to $\boldsymbol{X}_i = f_p(l_i)$, where $\mid \boldsymbol{X}_i\mid = N_p$.
To label the extracted packets for the use of later classification training, we compared features from the PCAP files to ground truth flow-level data provided by each testing dataset. 
An identifier feature \emph{Flow ID} is established for each flow to give corresponding packets the connection statistics, facilitating subsequent packet joint embedding generation.

\textbf{Compact Packet Feature Generation: } 
After obtaining packets $p_1,\dots,p_N$ with $N_p$ dimension transformed payload features $\boldsymbol{X}_i = [x^{1}_{i}, \dots, x^{N_p}_{i}]$, we train an autoencoder to learn a compressed representation of the payload features for each packet $p_i$. 
 Let $\boldsymbol{X}=[\boldsymbol{X}_1, \dots, \boldsymbol{X}_N] \in \mathbb{R}^{N \times N_p}$ be the input payloads from all packets $p_1,\dots,p_N$, where $N$ is the number of payload samples and $N_p$ is the number of payload features per sample. 
 The autoencoder consists of an encoder network, a decoder network, and a bottleneck layer with $N_b$ neurons, which is the number of payload features after compression. 
 The encoder network is defined as:
 \begin{equation} \label{eq:enc}
     \begin{aligned}
         \boldsymbol{H} = \sigma(W_1 \boldsymbol{X} + b_1),\ \ \boldsymbol{Z} = \sigma(W_2 \boldsymbol{H} + b_2),
     \end{aligned}
 \end{equation}
where $W_1 \in \mathbb{R}^{N_p \times h}$ and $W_2 \in \mathbb{R}^{h \times N_b}$ are the weight matrices of the encoder network, $b_1 \in \mathbb{R}^{h}$ and $b_2 \in \mathbb{R}^{N_b}$ are the bias vectors, $\sigma$ is the activation function (e.g., ReLU or sigmoid), and $\boldsymbol{H}$ is the hidden layer with $h$ neurons. The output of the encoder network is the compressed payload representation $\boldsymbol{Z} \in \mathbb{R}^{N_b}$.
And the decoder network is defined as:
\begin{equation} \label{eq:dec}
     \begin{aligned}
         \boldsymbol{H}' = \sigma(W_3 z + b_3),\ \ \boldsymbol{X}' = \sigma(W_4 \boldsymbol{H}' + b_4),
     \end{aligned}
 \end{equation}
where $W_3 \in \mathbb{R}^{N_b \times h}$ and $W_4 \in \mathbb{R}^{h \times N_p}$ are the weight matrices of the decoder network, $b_3 \in \mathbb{R}^{h}$ and $b_4 \in \mathbb{R}^{N_p}$ are the bias vectors, $\sigma$ is the activation function, and $\boldsymbol{H}'$ is the hidden layer with $h$ neurons. The output of the decoder network is the reconstructed input data $\boldsymbol{X}' \in \mathbb{R}^{N \times N_p}$.
The objective of the autoencoder is to minimize the reconstruction error between the input data $X$ and the reconstructed output data $X'$ using a loss function, here we use the mean squared error:
\begin{equation} \label{eq:ae_loss}
 L = \frac{1}{n} \sum_{k=1}^n | \boldsymbol{X}_k - \boldsymbol{X}_k'|^2,
\end{equation}
where $N$ is the number of payload samples, $\boldsymbol{X}_k$ is the original payload input data for the $k$-th sample, $\boldsymbol{X}_k'$ is the reconstructed output payload data for the $k$-th sample, and $|\cdot|$ denotes the L2-norm. 
In this case, once the autoencoder is fit, the reconstruction aspect of the model can be discarded and the model up to the point of the bottleneck can be used. The output $\boldsymbol{Z} \in \mathbb{R}^{N_b}$ of the model at the bottleneck is a fixed-length vector that provides a compressed representation of the input payload samples. Therefore, after this step, we obtain the sequence of the packets $\xi = [p_1,\dots,p_N]$ with compressed payload features $\boldsymbol{Z}_i = [z^{1}_{i}, \dots, z^{N_b}_{i}]$ for each packet $p_i$.

\textbf{Generating Packets Joint Embeddings: } 
Subsequently, we employ a \textbf{second} Recurrent Autoencoder to aggregate the packet-level features. This process contains the following four steps: (\romannumeral1) Selecting a representative sample of packet pairs from the entire packet dataset; (\romannumeral2) Training an Autoencoder to combine the payload features of the selected packet pairs; 
(\romannumeral3) With the trained encoder, adopting a greedy method that uses a hierarchical approach to recursively combine pairwise packet pairs corresponding to the same flow.

In detail, we employ the use of an Autoencoder for the combination of compressed payload features from multiple packets with the same flow ID, resulting in a flow embedding for each flow. 
We implement a Recursive Autoencoder (RAE) by applying Autoencoders recursively to a sequence of terms. Let $\boldsymbol{Z}_1, \boldsymbol{Z}_2 \in \mathbb{R}^{N N_b}$, represent the vector embeddings of two different payload features that have been compressed using the encoder of the first autoencoder. During the encoding phase, the composed vector embedding $\hat{\boldsymbol{Z}}(\boldsymbol{Z}_1,\boldsymbol{Z}_2)$ is calculated as:
\begin{equation}
    \begin{aligned}
    \hat{\boldsymbol{Z}}(\boldsymbol{Z}_1,\boldsymbol{Z}_2)= f(W_1[\boldsymbol{Z}_1;\boldsymbol{Z}_2]+b_1).
    \end{aligned}
\end{equation}
where $[\boldsymbol{Z}_1;\boldsymbol{Z}_2] \in \mathrm{R}^{N N_b}$ is the concatenation of $\boldsymbol{Z}_1,\boldsymbol{Z}_2$. $W_1\in \mathrm{R}^{N N_b\times 2N N_b}$ is the parameter matrix in encode phase, and $b_1\in \mathrm{R}^{N N_b}$ is the offset. Similar to RNN, $f$ again is a nonlinear function, e.g., tanh. In the decode phase, we need to assess if $\hat{\boldsymbol{Z}}(\boldsymbol{Z}_1,\boldsymbol{Z}_2)$ is well learned by the network to represent the composed terms. Thus, we reconstruct the term embeddings by:
\begin{equation}
    \begin{aligned}
    O(\boldsymbol{Z}_1;\boldsymbol{Z}_2)= g(W_2[\boldsymbol{Z}_1;\boldsymbol{Z}_2]+b_1).
    \end{aligned}
\end{equation}
where $O[\boldsymbol{Z}_1;\boldsymbol{Z}_2]$ is the reconstructed term embeddings, $W_2\in \mathrm{R}^{N N_b\times 2N N_b}$ is the parameter matrix for decode phase, and $b_2\in \mathrm{R}^{N N_b}$ is the offset for decode phase and function $g$ is another nonlinear function. For training purposes, the reconstruction error is used to measure how well we learned term vector embeddings. Let $\theta = \{W_1;W_2;b_1;b_2\}.$ We use the Euclidean distance between the inputs and reconstructed inputs to measure reconstruction error, i.e., 
\begin{equation}
    \begin{aligned}
    \mathrm{E}([\boldsymbol{Z}_1,\boldsymbol{Z}_2];\theta) = ||[\boldsymbol{Z}_1;\boldsymbol{Z}_2]-O[\boldsymbol{Z}_1;\boldsymbol{Z}_2]||_2^2.
    \end{aligned}
\end{equation}
For a given flow with multiple packets, we adopt a greedy method~\cite{white2016deep} to train our RAE and recursively combine pairwise payload embeddings. The greedy method uses a hierarchical approach – it first combines payload features of adjacent packets in each flow and then combines the results from a sequence of packets in an execution path. Fig.~\ref{fig:sys_diag} shows an example of how to combine the payload embeddings to generate a flow embedding. It shows a (binary) execution path with a sequence of four packets. The greedy method is illustrated as a binary tree. Node 1 gives the vector embedding for the first two packets' payload features combined payload = ($\boldsymbol{Z}_1 \% \boldsymbol{Z}_2$) encoded from terms $[\boldsymbol{Z}_1 \% \boldsymbol{Z}_2]$. Then, we continue to process the remaining instructions, e.g., Nodes 2 and 3, until we derive the final vector embedding (i.e., the flow embedding) for the packet sequences of the given execution path. 

For each sequence of packets $\xi= [p_1,\dots,p_N]$, where each packet $p_i$ has compressed payload features $\boldsymbol{Z}_i = [z^{1}_{i}, \dots, z^{N_b}_{i}]$, the obtained packet joint embeddings becomes $\hat{\boldsymbol{Z}}_i = [\hat{z}^{1}_{i}, \dots, \hat{z}^{N_b}_{i}]$. Note that the length of each group of packets $\mid \xi \mid$ is different. Therefore, for testing dataset $D$, we got the packet joint embeddings $\hat{\boldsymbol{Z}} = [\hat{\boldsymbol{Z}}_1,\dots,\hat{\boldsymbol{Z}}_N] \in \mathbb{R}^{N \times N_b}$ datasets for training DNN classifier. 



\subsection{Training a DNN-based policy and conversion into Decision Trees}



After we got the packet joint embeddings $\hat{\boldsymbol{Z}} = [\hat{\boldsymbol{Z}}_1,\dots,\hat{\boldsymbol{Z}}_N]$ for the testing dataset $D$, we use $\hat{\boldsymbol{Z}}$ as the input to the DNN to perform a binary- or multi-class classification.
We use the cross-entropy loss function which measures the dissimilarity between the predicted probability distribution and the true distribution of class labels, i.e.,
\begin{equation}
\label{eq:dnn_loss}
    J(\theta) = \frac{1}{N}\sum_{i=1}^{N}L(\hat{\boldsymbol{Z}}_i, y_i), \ L(\hat{\boldsymbol{Z}}, y) = -\sum_{k=1}^{K} y_k \log(\hat{y}_k).
\end{equation}
where $\theta$ represents the parameters of the DNN classifier, $N$ is the number of total training samples, $(\hat{\boldsymbol{Z}}, y)$ is the cross-entropy loss for a single input example, $K$ is the number of classes, $\hat{\boldsymbol{Z}}$ is the joint embedding and $y$ is the corresponding ground-truth label for the corresponding joint embedding, $y_k$ is the indicator function for class $k$ and $\hat{y}_k$ is the predicted probability of class $k$ from the DNN classifier.

We then transform the learned DNN-based policy parameterized by $\theta$ into a decision tree policy $\mathcal{T}$, which is inherently explainable. This conversion $F_T$ is achieved through the use of a teacher-student training process as described in \cite{7}, where the DNN acts as the teacher and generates input-output samples to train the decision tree student. 
In detail, the teacher DNN $\theta$ first generates the label-predicted sequence of packet joint embeddings $(\hat{\boldsymbol{Z}},\hat{y})$, where $\hat{y}$ is the predicted label for joint embedding $\hat{\boldsymbol{Z}}$ trained by DNN classifier $\theta$. This generated data is then used as input to train the decision tree policy $\mathcal{T} = F_{\rm T} (\theta, \alpha)$, where the function $F_{\rm T}(\cdot)$ denotes the Classification and Regression Tree (CART) algorithm~\cite{cart}, and $\alpha$ denote the pruning parameters, i.e., `ccp\_alpha' to adjust the tree sizes.

\subsection{Modeling a Memristor-based Implementation }
The trained decision tree policy $\mathcal{T} = F_{\rm T} (\theta, \alpha)$ could benefit from an efficient hardware implementation, particularly in resource-constrained environments. We propose an analog front-end design $ \mathcal{H} = F_{\rm H} (\mathcal{T}, \beta)$ with each adjustable boundary implemented as a memristor device. This enables the programmed boundary to be compared fast via hardware acceleration with an input stream from the dataset. We quantize the splitting thresholds for each node of the trained tree $\mathcal{T}$ using $\beta$ bits, which indicate $2^{\beta}$ quantization levels.

\begin{figure}[b]
\centering
\includegraphics[width=3.55in, height=1.5in]{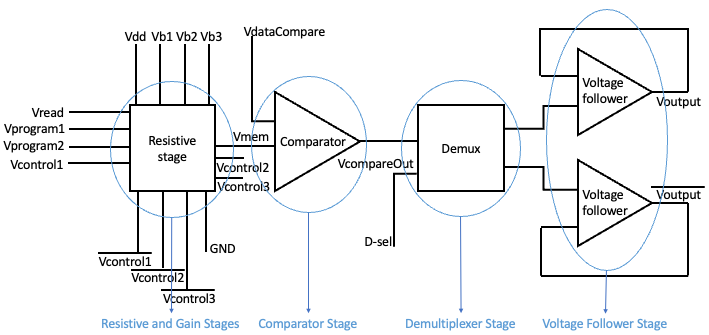}
\caption{Diagram of the memristor implementation design for the single node of a pruned decision tree.}
\label{fig:circuit_single_node}
\end{figure}

The hardware implementation of the decision tree $\mathcal{T}$ is envisioned as a distributed network based on small chiplets as leaf nodes. All single nodes in the design are identical for reusability and to reduce the design cost. 
The design of the node in the decision tree $\mathcal{T}$, consists of 5 stages supporting the signal pipeline shown in Fig.~\ref{fig:circuit_single_node}. The 5 stages are the resistive stage, the gain stage, the comparator stage, the demultiplexer stage, and the voltage follower stage. 130nm CMOS and memristor hybrid technology were used for the design and simulations of these stages.

\subsection{Co-design and Joint Optimization for software and hardware}


For the \textit{JSHC} optimization problem defined in Eq.~\ref{eq:0}, now we have $F_T$ in Eq.~\ref{eq:F_T} and $F_H$ in Eq.~\ref{eq:F_H}. Since the JSHC problem has the hardware constraint $ \zeta (\mathcal{H}) \le Z_{\rm max}$ which is not a closed-form function, we need to consider the trade-offs between the learning accuracy and the feasibility to deploy the learning architecture $F_T$ in the $F_H$ hardware environment. 

For adjusting the software architecture $F_T$, we employ the cost complexity pruning (CCP) technique\cite{26} to reduce the size of the decision tree with `ccp\_alpha' parameter $\alpha$ for hardware implementation. During this process, we fix hardware parameter $\beta$ in the optimization problem defined in Eq.~\ref{eq:0} without considering circuit constraints $\zeta (\mathcal{H}) \le Z_{\rm max}$.
The cost-complexity pruning is an algorithm used to prune a tree to avoid over-fitting. This algorithm is parameterized by $\alpha \geq 0$ known as the cost-complexity pruning alpha parameter, i.e., `ccp\_alpha' as mentioned above. The `ccp\_alpha' is used to define the cost-complexity measure, $R_{\alpha}(\mathcal{T})$ of a given decision tree $\mathcal{T}$:
\begin{equation}
    R_{\alpha}(\mathcal{T}) = R(\mathcal{T}) + \alpha \mid\Tilde{\mathcal{T}}\mid ,
\end{equation}
where $\mid \Tilde{\mathcal{T}} \mid$ is the number of terminal nodes in $\mathcal{T}$ and $R(\mathcal{T})$ is traditionally defined as the total misclassification rate of the terminal nodes. Alternatively, we use the total sample weighted impurity of the terminal nodes for $R(\mathcal{T})$. As shown above, the impurity of a node depends on the criterion. CCP algorithm finds the subtree of $\mathcal{T}$ 
 that minimizes $R_{\alpha}(\mathcal{T})$.
The cost complexity measure of a single node is $R_{\alpha}(t)=R(t)+\alpha$. The branch $\mathcal{T}_{t}$ is defined to be a tree where node $t$ is its root. In general, the impurity of a node is greater than the sum of impurities of its terminal nodes, i.e., $R(\mathcal{T}_t) < R(t)$. However, the cost complexity measure of a node $t$ and its branch $\mathcal{T}_t$ can be equal depending on $\alpha$. We define the effective $\alpha_{eff}$ of a node to be the value where they are equal:
\begin{equation}
    R_{\alpha_{eff}}(\mathcal{T}_t) = R_{\alpha_{eff}}(t), \ \  or \ \ 
    \alpha_{eff}(t) = \frac{R(t) - R(\mathcal{T}_t)}{\mid \mathcal{T}\mid-1}.
\end{equation}
A non-terminal node with the smallest value of $\alpha_{eff}$ is the weakest link and will be pruned. This process stops when the pruned tree’s minimal $\alpha_{eff}$ is greater than the `ccp\_alpha' which is the input  $\alpha$ in Eq.~\ref{eq:F_T}. 
In order to find the optimal $\alpha$, we fix the hardware parameter $\beta$ and use an exhaustive search process - every $\alpha$ in a pre-defined set is checked before a decision of the optimal $\alpha$ is made. Thus, the cost complexity pruning results in a more compact tree that can then be implemented using a flexible memristor architecture. 

\begin{figure}[t]
\centering
\includegraphics[width=2.4in, height=1.8in]{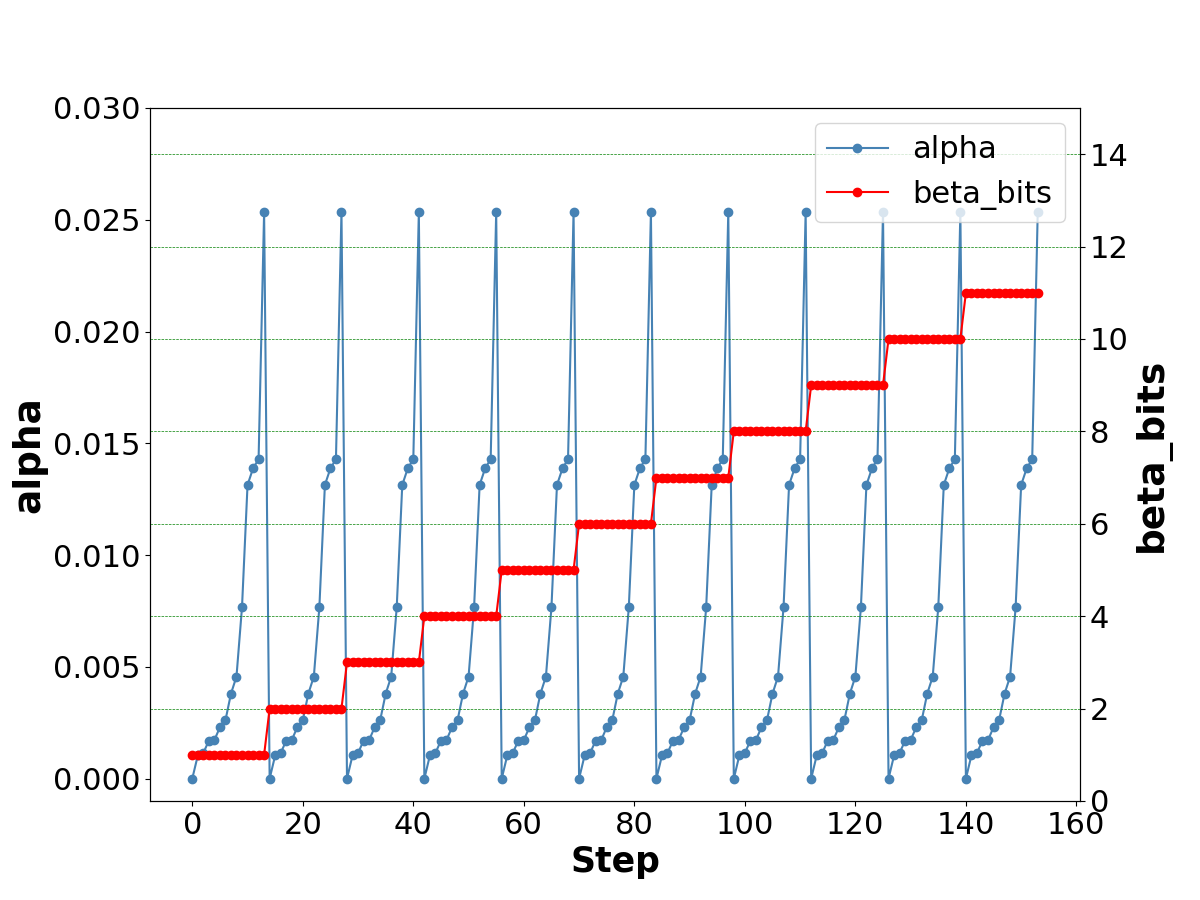}
\caption{Iteratively tuning one parameter, $\alpha$ (decision tree pruning parameter), or $\beta$ (number of bits), while keeping the other fixed during joint optimization.}
\label{fig:alpha_beta_adjust_process} 
\end{figure}

From a hardware design perspective, a higher quantization level $2^{\beta}$ corresponding to a larger number of bits $\beta$ is used, which involves more circuit complexity as well as additional power and area. However, in order to make power more efficient and the chipset smaller, the optimization of the trade-off is needed. 
In order to find the optimal memristor-based circuit design which is parameterized by the number of bits $\beta$ used for quantizing the splitting thresholds at each node of the trained decision tree $\mathcal{T}$, we fix the `ccp\_alpha' parameter $\alpha$ from the software side and use a bisection search to find the optimal $\beta$. The step-by-step tuning process of $\alpha$ and $\beta$, where one parameter is tuned while keeping the other fixed, is visualized in Fig.~\ref{fig:alpha_beta_adjust_process}.

\section{Computational Experimentation}

\subsection{Datasets}

The availability of well-labeled datasets with real network traffic is a challenge for payload-based intrusion detection research, as many datasets used for network intrusion detection only comprise header information of network packets or lack payload data due to privacy considerations~\cite{hassan2021intrusion}. To select a suitable dataset to evaluate our algorithm, we reviewed available network intrusion detection datasets that meet our established criteria for the presence of flow-based and packet-based data, anomalous events, and labeled payloads. We selected two datasets that meet our criteria, namely UNSW-NB15~\cite{unsw} and CIC-IDS-2017~\cite{cicids}, both of which contain payload data in packets and flow data.

\subsection{Experiment Results}
First, we evaluate the effect of compressed dimension $N_b$ during the process `Compact Packet Feature Generation' in our proposed algorithm in Section.~\ref{subsec:pac_joint_embeddings_generation} on the UNSW-NB15 and CIC-IDS-2017 datasets for binary and multi-class intrusion detection. This part uses an autoencoder to compress payload features to be more compact. We conduct an experiment to compare the testing accuracy scores attained by compressing input payload features into various shapes $N_b$ using the UNSW dataset, which calculates the accuracy of a logistic regression classification model by comparing the true labels with the predicted labels on the testing dataset. The results in Table~\ref{table:payload_compress} demonstrate that compressing payload features into various shapes does not significantly affect detection accuracy, so we compress the payload features from $N_p=$1500 dimensions to $N_b=$100 dimensions in the subsequent experiments to handle the lower dimension of the payload feature. 

\begin{figure}[t]
\centering
\includegraphics[width=3.2in, height=1.6in]{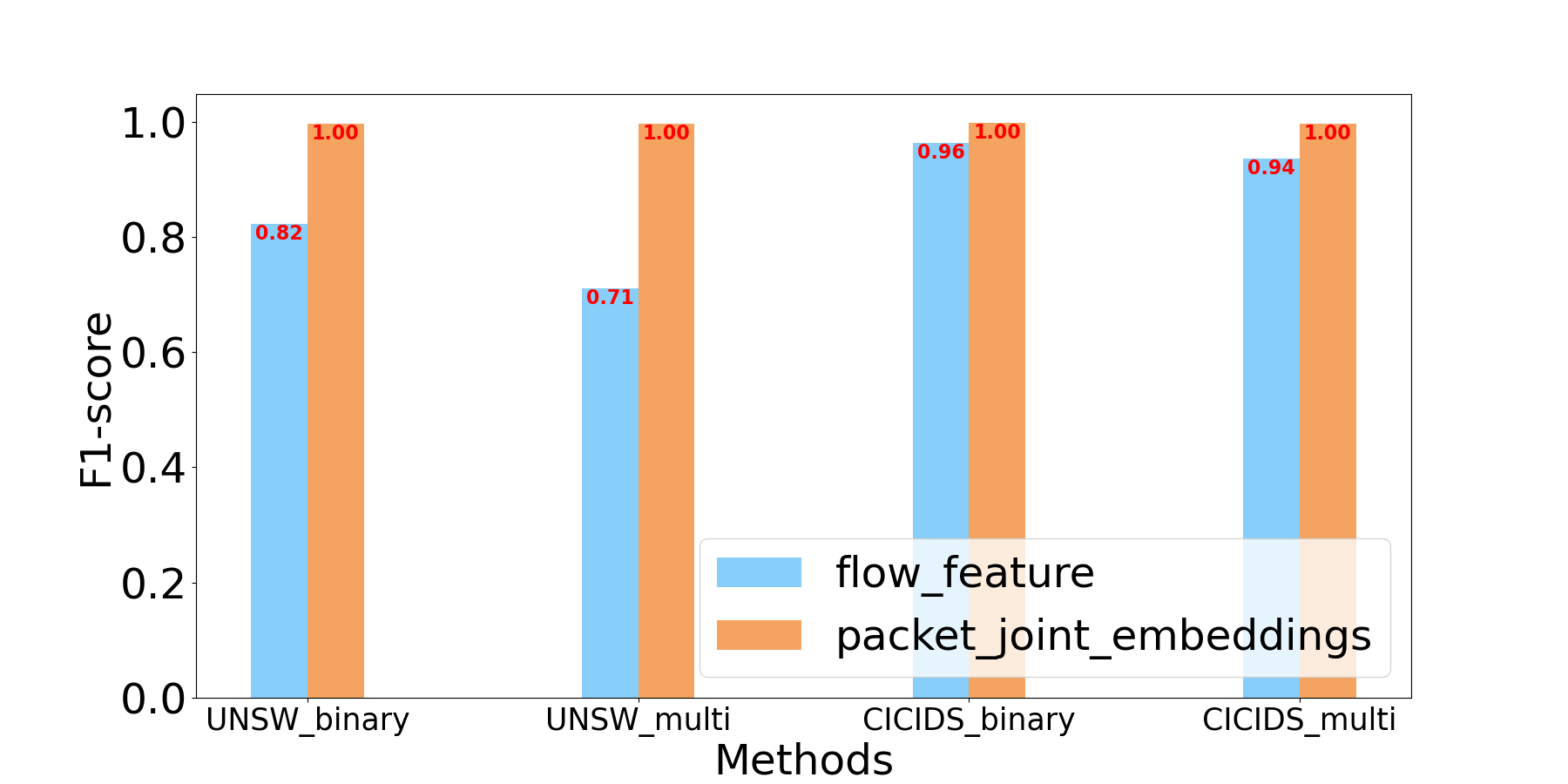}
\caption{Packet joint embeddings (orange bars) outperform flow-level features (blue bars) in Intrusion Detection, achieving higher F1-scores in binary- and multi-class classification for the UNSW-NB15 and CIC-IDS-2017 datasets.}
\label{fig:set_0_fig} 
\end{figure}

After extracting and compressing payload features, we compare the intrusion detection accuracy of binary and multi-class classification using the packet joint embeddings generated from the UNSW-NB15 and CIC-IDS-2017 datasets against their original flow-level datasets. The results in Fig. \ref{fig:set_0_fig} show that using a Recurrent Autoencoder to compress and combine packet features into a more compact joint embedding can significantly improve detection accuracy. This suggests that packet joint embeddings that incorporate payload information from multiple packets contain more efficient information for network intrusion detection.

\begin{table*}[htbp] 
  \begin{center}
    \caption{Comparing the testing accuracy scores of compressing packet payload features to different shapes.}
    \label{table:payload_compress}
    \begin{tabular}{|c|c|c|} 
    \hline
      \textbf{Compressed Payload Shape} & \textbf{Accuracy (Binary-Class)} & \textbf{Accuracy (Multi-Class)}\\
      \hline
     1500 & 97.88\% & 98.33 \%\\
     \hline
     750 & 97.42\% & 96.36\% \\
     \hline
     500 & 96.97\% & 96.81\% \\
     \hline
     300 & 97.73\% & 97.87\%\\
     \hline
     100 & 97.57\% & 99.39\% \\
     \hline
    \end{tabular}
  \end{center}
\end{table*}

Then we compare the classification accuracy using F1 scores, inference time, and learning model size of the DNN classifier, Decision Tree without pruning, and Pruned Decision Tree in memristor-based implementation in Table~\ref{table:total_compare}. Our proposed co-design approach using packet joint embeddings (\textit{UNSW: Pruned Decision Tree with packet joint embeddings (memristor)} and \textit{CIC-IDS-2017: Pruned Decision Tree with packet joint embeddings (memristor)}) achieves nearly four orders of magnitude speed up while retaining the desired accuracy, it shows pruning and implementing the decision tree that is extracted from DNN policy on memristor provides human-readable interpretations while preserving nearly no degradation in performance and further allow the computing system to have higher density and to be more energy efficient due to characteristics of our memristor-based circuit.

\begin{table*}[htbp] 
  \begin{center}
    \caption{Comparisons with baselines on UNSW-NB15 and CIC-IDS-2017 dataset}
    \label{table:total_compare}
    \begin{tabular}{|c|c|c|c|} 
    \hline
      \textbf{Method} & \textbf{F1 score} & \textbf{Inference time(s)} &  \textbf{Learning model size}\\
      \hline
       UNSW: MLPClassifier with packet joint embeddings & $\boldsymbol{0.9962}$ & 76.7515 &1 layer/100 neurons\\
      \hline
 UNSW: Decision Tree with packet joint embeddings & 0.9817 & 0.0106 & 35 nodes\\
      \hline
 UNSW: Pruned  Decision Tree with packet joint embeddings (memristor) & 0.9771 &$\boldsymbol{0.00004}$& \textbf{15 nodes}\\
\hline
    CIC-IDS-2017: MLPClassifier with packet joint embeddings & $\boldsymbol{0.9980}$ & 20.7829 &1 layer/100 neurons\\
      \hline
     CIC-IDS-2017:  Decision Tree with packet joint embeddings & 0.9973 & 0.9707 & 23 nodes\\
      \hline
 CIC-IDS-2017: Pruned  Decision Tree with packet joint embeddings (memristor) &  0.9966 &$\boldsymbol{0.00002}$& \textbf{5 nodes}\\
\hline
    \end{tabular}
  \end{center}
\end{table*}

Next, we conduct an analysis of our proposed algorithm to evaluate the impact of varying parameter values on metrics from both the software and hardware side, i.e., detection accuracy, inference time, circuit additional area consumption, and power consumption. Specifically, we focus on the UNSW-NB15 binary-class classification scenario and employ our algorithm, which is designed for the co-optimization of both software and hardware.
\begin{figure}[htbp]
\centering
\subfigure[Test Scores]{
\label{fig:no_beta_alpha_acc_nodes} 
\includegraphics[width=1.65in, height=1.95in]{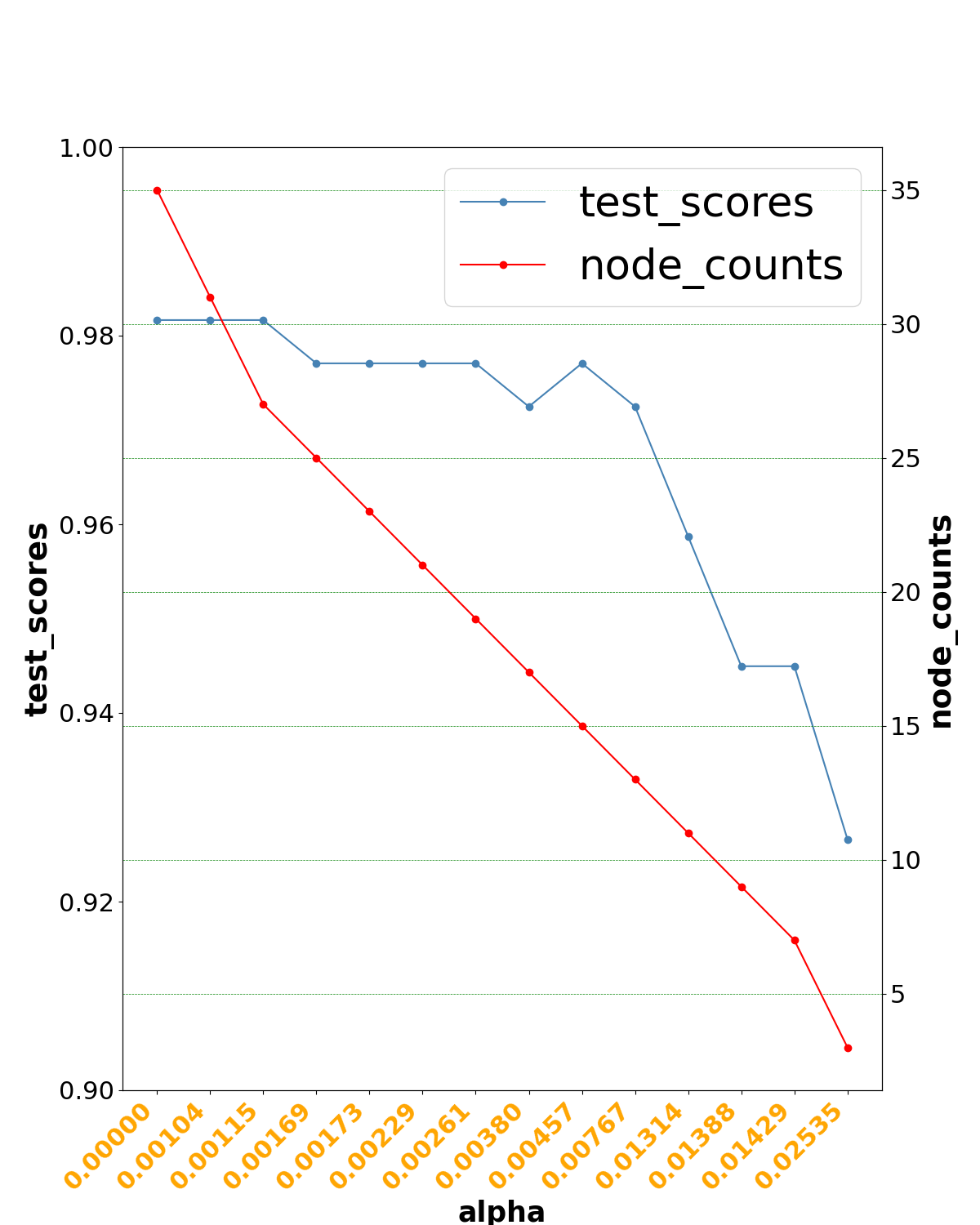}}
\subfigure[Inference Time]{
\label{fig:no_beta_alpha_InferenceTime_nodes} 
\includegraphics[width=1.65in, height=1.95in]{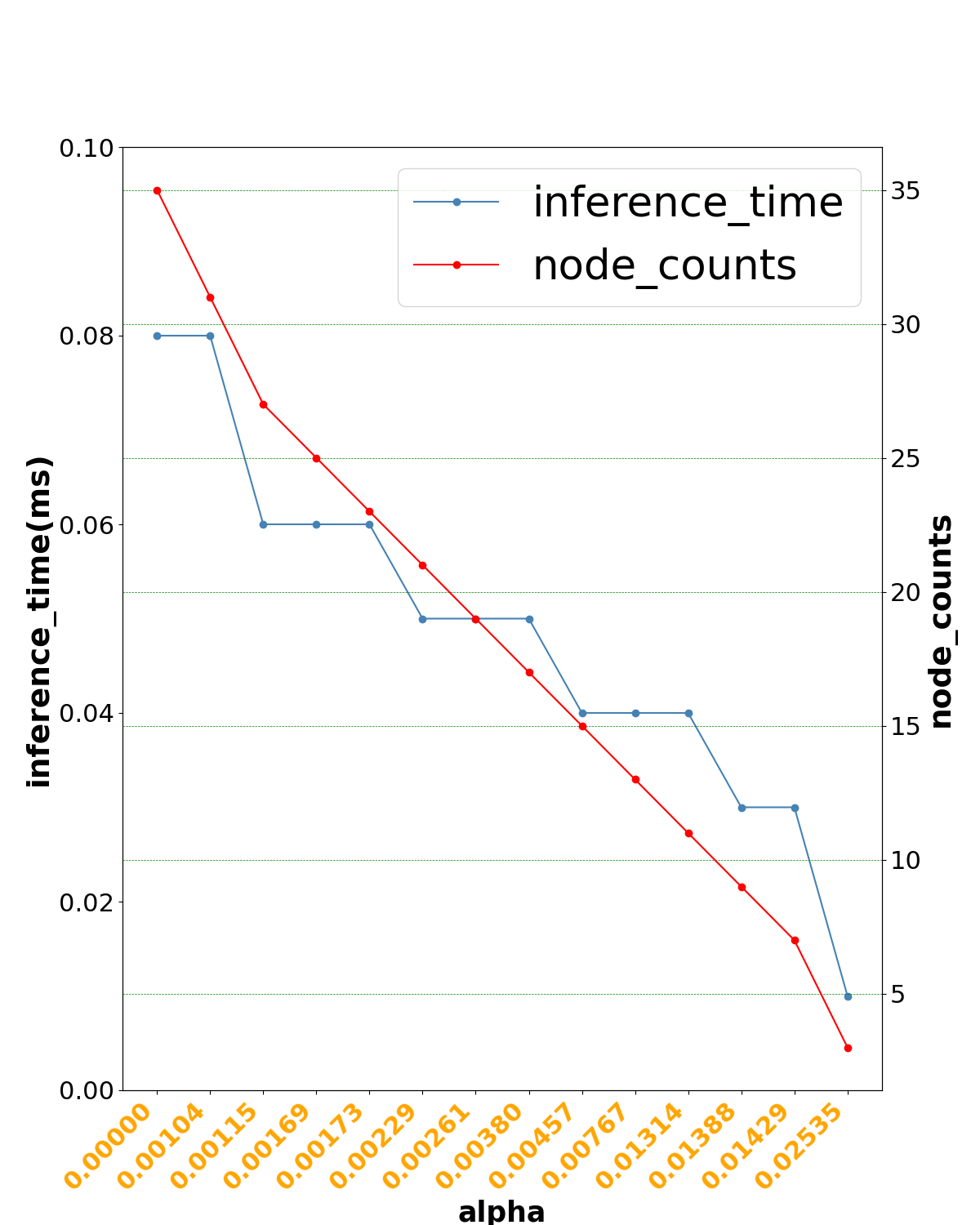}}
\caption{As the value of decision tree pruning parameter `ccp\_alpha' $\alpha$ increases, the number of nodes in the tree decreases. This reduction in complexity causes a decrease in learning accuracy and in inference time.}
\label{fig:no_beta_alpha_acc_time_nodes} 
\vspace{-.1in}
\end{figure}


\begin{figure*}[htbp]
\centering
\subfigure[Testing scores]{
\label{fig:beta_acc} 
\includegraphics[width=2.3in, height=1.5in]{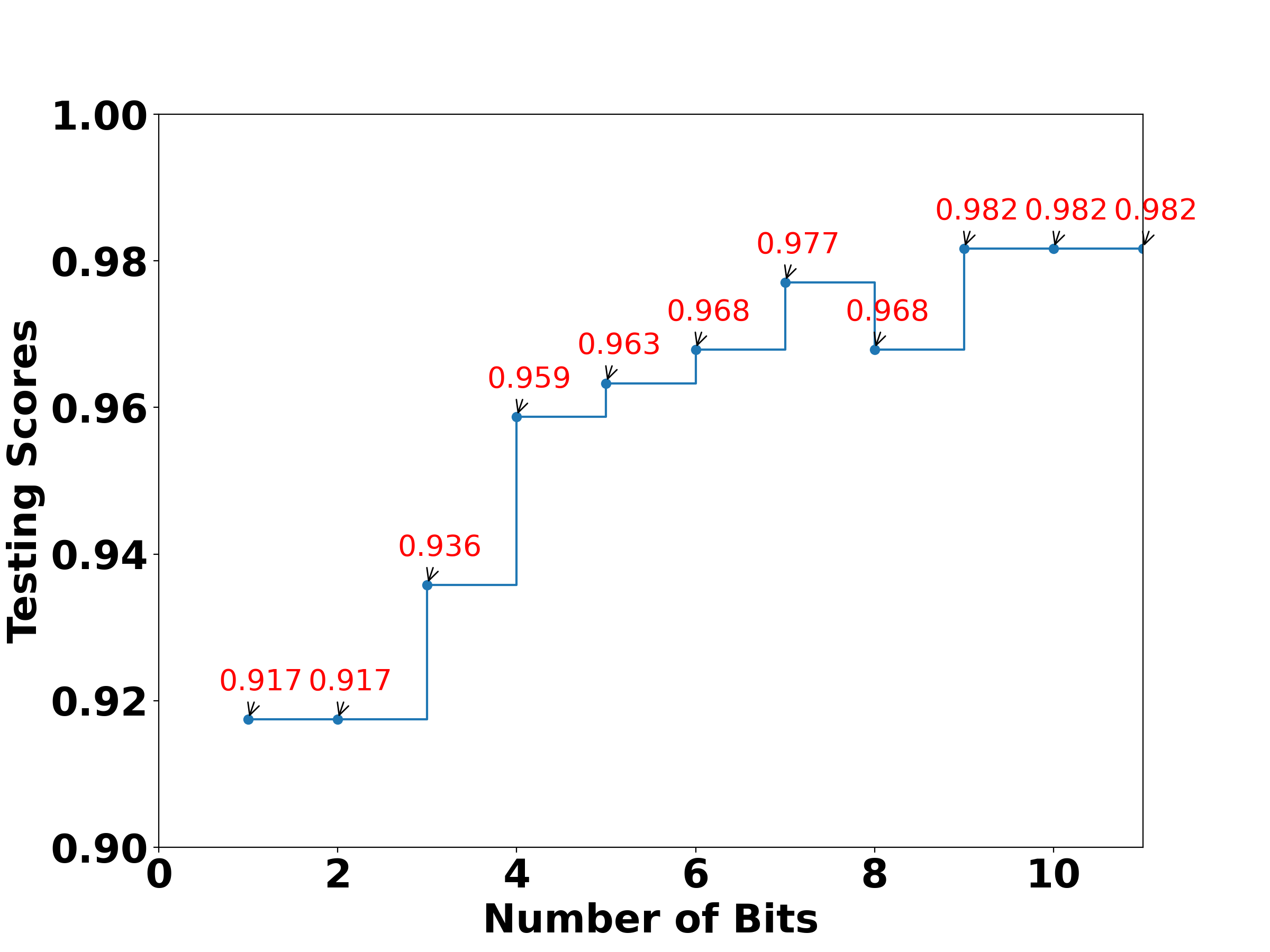}}
\subfigure[Power consumption]{
\label{fig:beta_power} 
\includegraphics[width=2.3in, height=1.5in]{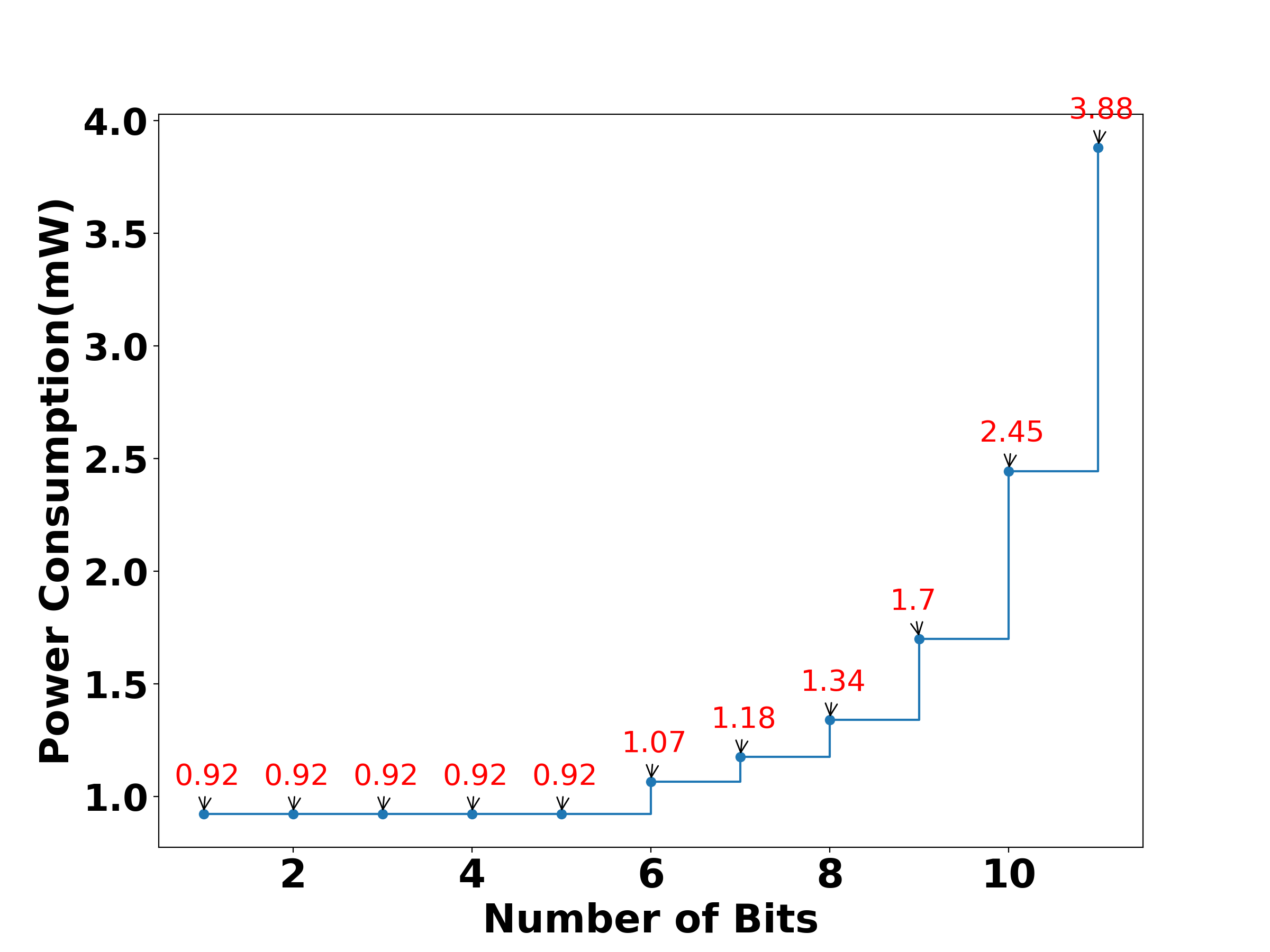}}
\subfigure[Additional Area Consumption]{
\label{fig:beta_area} 
\includegraphics[width=2.3in, height=1.5in]{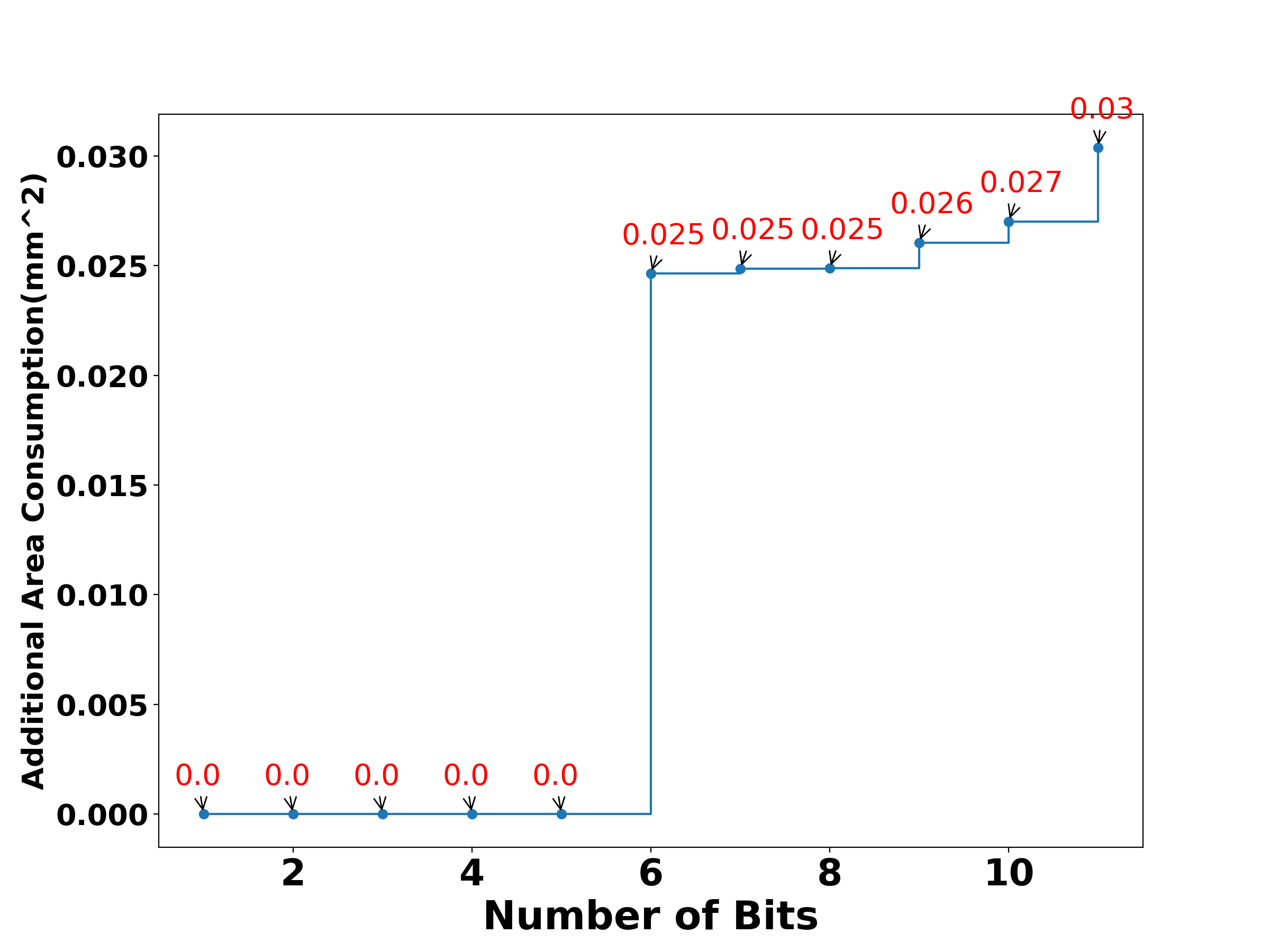}}
\caption{Setting $\alpha=0$ for decision tree pruning, increasing the number of bits $\beta$ improves testing scores but increases power and area consumption, highlighting the need to balance accuracy and resource usage.}
\label{fig:beta} 
\vspace{-.1in}
\end{figure*}

\begin{figure*}[h]
\centering
\subfigure[Accuracy]{
\label{fig:contour_acc} 
\includegraphics[width=2.3in, height=1.5in]{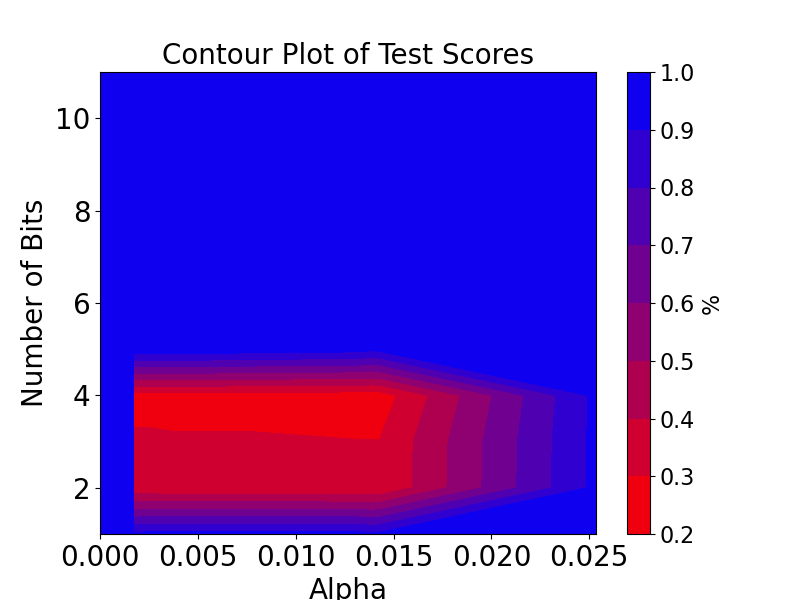}}
\subfigure[Power consumption]{
\label{fig:contour_power} 
\includegraphics[width=2.3in, height=1.5in]{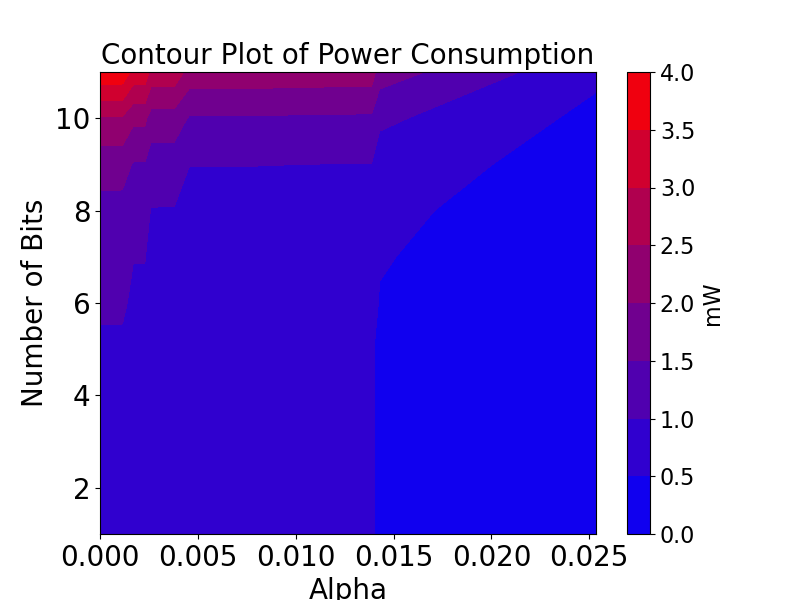}}
\subfigure[Addition Area Overhead]{
\label{fig:contour_area} 
\includegraphics[width=2.3in, height=1.5in]{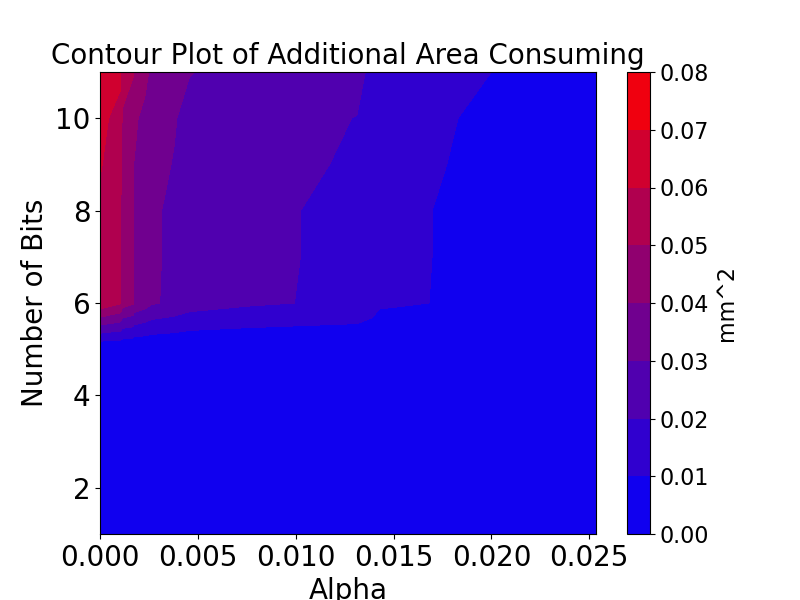}}
\caption{The contour plot shows the joint optimization of $\alpha$ (decision tree pruning parameter) and $\beta$ (number of bits). A smaller $\alpha$ and larger $\beta$ yield high accuracy but result in higher power consumption and additional area overhead.}
\label{fig:alpha_beta_combine} 
\vspace{-.1in}
\end{figure*}


We begin by analyzing the impact of varying `ccp\_alpha'(represented by $\alpha$) on learning accuracy and inference time while fixing the number of bits of quantization levels of decision tree's boundaries (represented by$\beta$) in Fig.~\ref{fig:no_beta_alpha_acc_time_nodes}. The decision tree extracted without pruning has over 30 nodes. In Fig.~\ref{fig:no_beta_alpha_acc_nodes}, as $\alpha$ increases, both the number of nodes and learning accuracy decrease. For this scenario, setting $\alpha$ to 0.0026 maximizes testing accuracy at $0.97706422$ and yields a decision tree with 19 nodes that can be implemented more efficiently in hardware.
Furthermore, Fig.~\ref{fig:no_beta_alpha_InferenceTime_nodes} shows that a larger value of $\alpha$ leads to a smaller decision tree with fewer nodes and thus less inference time when implemented in hardware.

To investigate the impact of the hardware parameter $\beta$, we set $\alpha$ to $0.0000$ representing a decision tree without any pruning, and vary the value of $\beta$ in Fig.~\ref{fig:beta}. Fig.~\ref{fig:beta_acc} shows that when fixing the alpha, the learning accuracy will increase from $0.917$ and converge to $0.982$ as $\beta$ increases from $1$ to $11$, which indicates that a larger number of bits $\beta$ could lead to a higher accuracy since it relates to a more complex circuit, which is shown in Fig.~\ref{fig:beta_power} and Fig.~\ref{fig:beta_area} - as the number of bits $\beta$ increases, the power consumption and additional area consumption will both increase. 
Fig.~\ref{fig:beta_power} and Fig.~\ref{fig:beta_area} indicate that the complexity of the circuit will increase when the $\beta$ is $6$ or above. To support a maximum $11$-bit level, $2^{11}=2048$ numerical levels are needed, which requires a large output voltage swing. The additional amplifier and input/output pins have to be added to the simulation of the circuit for each bit level $\beta$ above $6$. From the comprehensive simulation results of the memristor hardware implementation, when the numerical precision of the boundary is at $11$ bit level, the power consumption is $3.88 mW$ for the whole $35-$node decision tree for inference, which is four times larger than the $5$ bits or less, but in the meantime, the precision has a $64$ times enhancement. The difference in the silicon area consumed across the levels does not change much. The inference time does not change a lot while the precision is increasing. 


The impact of different combinations of $\alpha$ and $\beta$ on the learning accuracy and inference time is presented in Fig.~\ref{fig:alpha_beta_combine}. 
These figures display contour plots to show the joint optimization of $\alpha$ and $\beta$. From Fig.~\ref{fig:contour_acc}, it is evident that maximizing the learning accuracy requires a larger number of bits and a smaller pruning parameter. This is because a larger $\beta$ corresponds to a more complex circuit, and a smaller $\alpha$ corresponds to a decision tree with more nodes. 
Fig.~\ref{fig:contour_power} and Fig.~\ref{fig:contour_area} show that using less number of bits to implement a smaller decision tree (a larger $\alpha$) will have less power consumption and additional area consumption in the circuits.

\section{Conclusions}
We propose a new approach for packet-level intrusion detection that integrates multiple packets with varying payload features into joint packet embeddings. Our method is designed for hardware mapping, suitable for constrained computational environments with high detection performance. We emphasize co-exploration of hardware and algorithmic technologies, leveraging cybersecurity, machine learning, and nanoscale device advances. The proposed packet joint embeddings extraction algorithm and the memristor hardware-mappable algorithm achieve impressive results: three-nines detection accuracy and a significant four orders of magnitude speedup, respectively. Future work will explore advanced decision architectures and experimental hardware prototypes.

\section*{Acknowledgments}
This work was supported in part by the U.S. Military Academy (USMA) under Cooperative Agreement No. W911NF-22-2-0089 and the U.S. Army DEVCOM Army Research Laboratory under Support Agreement No. USMA 21050. The views and conclusions expressed in this paper are those of the authors and do not reflect the official policy or position of the U.S. Military Academy, U.S. Army, U.S. Department of Defense, or U.S. Government. 


\bibliographystyle{plain}
\bibliography{main_IEEESec}

\end{document}